\documentclass[aps, showpacs]{revtex4}
\begin{document}
\title{On the Dark Energy}
\author{S.C. Tiwari,
Institute of Natural Philosophy,
Varanasi 221005, India; Present affiliation: Department of Physics, Institute of Science, Banaras Hindu University, Varanasi 221005, India}
\begin{abstract}

The dark energy problem may have a simple solution in the model of cosmology based on the space-time interaction hypothesis. The hypothesis throws light on the nature of
time (see Time-Transcendence-Truth, Monograph published in 2006).
\end{abstract}
\maketitle

\section{\bf Introduction}

Ramifications of recent observations on the anisotropy and fluctuations in the cosmic
microwave background (CMB) radiation, and magnitude versus redshift of type Ia Supernovae (SN Ia) on the history of cosmic evolution from the moment of big-bang creation
are being intensely explored. These observations strongly suggest a flat universe at large
scales in the standard model of cosmology based on general relativity (GR). Accelerating
expansion of the universe is also indicated. For vanishing large scale curvature, the total
average energy density in the universe has to be close to the critical density, $\rho_0=  10^{-29}$
gm/cm$^3$. Estimated mean mass density from the astronomical observations is, however less
than $5\%$ of this value. Note that this estimate refers to baryonic matter and radiations
including neutrinos. Even finite non-zero mass of neutrinos changes this value marginally.
We do not know what kind of $95\%$ of the energy in universe is though it has been proposed
that $25\%$ of $\rho_0$  is in the form of dark matter (DM) and rest is dark energy. Interestingly as
early as 1937 Zwicky had anticipated DM content, and later on cold DM models for the
structure of galaxies and clusters of galaxies were extensively studied. Temperature
fluctuations in the CMB data show that $\Omega_{DM} \approx 0.25 to 0.30$ where $\Omega_{DM} = \rho_{DM}/\rho_0$ .
Astrophysicists favor cold DM models, and recent X-ray observations \cite{1} seem to be
consistent with them. The best fits to the data in Hubble diagrams for SN Ia are obtained for
the mass density that implies that galaxies are moving away from each other at an accelerating
rate \cite{2}. Constraints on dark energy and its equation-of-state parameter have recently been
reported analyzing the observed data from high redshift supernovae and Hubble Space
Telescope \cite{3}. The most favored physical explanations for dark energy are cosmological
constant and quintessence, see the review \cite{4}. In spite of enormous activity on this subject, the
nature of dark energy remains mysterious and it is hoped that its resolution would profoundly
affect fundamental physics \cite{2, 5}.

In theoretical physics, quantum gravity (QG) has proved to be the biggest stumbling-block in the quest for unification. Superstrings and loop quantum gravity are believed to be
the leading theories in the mainstream physics, though canonical quantization is also being
pursued \cite{6}. Unlike dark energy problem, QG has essentially a philosophical motivation: if
both GR and quantum theory are fundamental, they ought to be unified. Another argument for
QG stems from the speculation that in the early universe, the classical picture of space-time
would fail during Planck epoch and quantization would become imperative. Note that Planck time is $\tau_P \approx 10^{-44}$sec and the corresponding length is $ 10^{-33}$cm. While from the observational point of view, early
universe implications are very uncertain, there are significant questions that can be raised at a
conceptual level. Is spacetime illusion? Is spacetime discrete? What is the meaning of the
arrow of time ? May be QG will lead to a revision of spacetime picture or may be there exists
an underlying radically different reality, e.g. spin networks \cite{7} and spin foams \cite{8} from which
emerges spacetime geometry. Carlip in \cite{6} makes a brief reference to such radical departures.
The aim of this short essay is to re-visit the space-time interaction hypothesis \cite{9} that
offers an elegant and simple resolution to the dark energy puzzle. 

\section{\bf New Approach}

Space-time interaction hypothesis has a potential for developing a new approach to the cosmological problems and throws light on the
nature of space and time. One of its speculative consequences, namely the time-varying speed
of light, has recently been discussed in the context of various scenarios of superluminal
phenomenon in \cite{10}. For the sake of self-contained discussion let us first state the postulates as presented in \cite{9}.

{\bf Postulate I} :
 There is a state ‘superspace’ which has its own energy called fundamental energy.
 
 {\bf Postulate II}:
 Time is something which operates upon the state ‘superspace’ to transform the
fundamental energy to known from of energy thereby creating universe.

Since the words like ‘superspace’ and ‘fundamental energy’ have different
connotations in the contemporary physics literature, we refer to the later expositions and
elucidations of this hypothesis to avoid confusion \cite{10-11}. Superspace or source space is a
pre-primordial state before the universe came into being, and the time is a cause of the
becoming of the universe. We may view the universe creation as a spontaneous phase
transition such that time is a primeval cause of this transition i.e. time unfolds the universe.
Though we associated energy with the superspace in the original statement of the hypothesis,
it is not necessary at an abstract level. We could simply say the quantity of space instead. A sequential, finite duration act of transformation is envisaged, with each succession of
transformation the quantity of manifest universe increases. It was argued that the Nature
should not be irregular and partial implying that the length of time duration in each act of
transformation is equal, and the quantity of space transformed is also equal. While writing
the monograph \cite{10}, I became aware of the original saying, cause acequat affection \cite{12}
and the interpretation in classical physics that the cause and effect are two successive forms of
the constant quantity of energy. It is interesting that this form of causality is inherent in our
hypothesis. The nature of time is radically different; it is discrete, unidirectional and action
itself. On the order hand, the space is a continuum in the sense of wholeness. How is this
universe related with the observable physical universe?

To answer this question we have to introduce certain assumptions and definitions. Let
us assume spherical geometry of space, elementary volume $V_e$ is created in each elementary
time interval assumed to be Planck time $\tau_P$. Planck constant is treated as a unit conversion
parameter, and introducing some energy conversion unit, $V_e$ can be equivalently expressed as
energy. Evolution of universe proceeds as expanding sphere, and at each act of transformation
spherical shells are created. Simple calculations \cite{10} show that beginning with a size of about $10^8$cm, within less than a second spatial regions of $\approx 10^{-21}$cm came into existence, and the length
scales of the order of Planck length are created at the age of about $10^10$ years. Contrast this with big-bang model in which very early universe, i. e. the Planck era, is believed to have this size. Let us denote such elementary spatio-temporal structures symbolically by $E_i , i = 1,2,... N$. Each $E_i$
is characterized by internal time, translational motion and spin. The proposed structures are
endowed with Gaussian-like shapes such that the tails make the whole universe a space
continuum. Two or more than two $E_i$  can combine to form lumps or aggregates; such spatio-temporal aggregates constitute matter. At large scale, in the history of universe, statistical
laws determine the distribution of $E_i$  with spatial sizes ranging from $10^8$cm down to the size
created at that age of the universe. Qualitative picture that emerges from these considerations is as follows.

Radial expansion rate of the universe is decelerating, and the radial expansion velocity
serves the purpose of a cosmic limiting velocity. Note that the velocity of the order of $10^22$cm/sec existed at the age of one second of the universe. Since the elementary time interval, $\tau_P$ 
or its integral multiples are fundamental in our model, the record of possible high speeds in
the past exists as a memory and could be recalled to exceed present day speed of light, i.e. the
limiting velocity principle is not absolute \cite{10}. In the early universe, $E_i$ with different sizes
and very high speeds were in random motion colliding with each other, forming aggregates
and so on, while the universe also was undergoing fast expansion. It is possible that as
universe evolved, a part of it attained almost thermodynamical equilibrium. We conjecture
that our solar system and galaxy are possibly located in that region i.e. the matter we see
around us became synthesized in that region. The region in the vicinity of the boundary of the
universe would be in a highly non-equilibrium state. Since the energy density averaged over
whole of the universe is always constant, we are free to assume it equal to $\rho_0$, i.e. there is no
dark energy puzzle. As regards to the accelerating phenomenon in SN Ia \cite{2}, a tentative
possibility is to imagine these objects in the vicinity of the boundary of the universe where
non-equilibrium state leads to formations and disintegrations of the structures that fly apart at an accelerating rate.

\section{\bf Fundamental Physics}

The basic building blocks of universe in our model are $E_i$. How does one construct matter from them? What are the fundamental interactions? The spatio-temporal structures are
fundamental, therefore we do not have to postulate quarks, preons, ad infinitum. The $E_i$
combine with each other by some geometrical rules, and spin-spin interaction is proposed to
be the fundamental interaction. Perhaps spin networks \cite{7} or some other combinational rules
will guide us to develop this idea. A modest, though radically new approach has been pursued
based on the assumption that neutrino(s), electron (positron) and photon are primary
elementary objects for matter \cite{13}. In this approach, electronic charge is given a mechanical
interpretation as a manifestation of fractional spin $e^2/c$ and the Maxwell field equations
represent rotational photon fluid.
Our hypothesis has some testable consequences: time-varying velocity of light, life-time of an unstable particle and non-equivalence of inertial frames, and those related with
photon-fluid \cite{10,13}. Alternative cosmological model proposed here can be put into rigorous
computation, e.g. by suitably changing the algorithm of \cite{14} or the ones used in variable
speed of light models \cite{10}. Such a calculation will result into the history of universe that can
be tested against CMB observations and the deduced values of 
$\Omega_B$ and $\Omega_{DM}$Ω DM.

The most significant contribution of our work is the insight gained on the nature of
space and time. A detailed critique on relativistic spacetime has been presented earlier \cite{10,11}; here we focus our brief remarks on some ideas that deny physical reality to space-time.
Penrose attempts to build spacetime from discrete angular momentum and combinational
principles. There seems to be a logical flaw since the concept of angular momentum presupposes space, see also a comment on p.148 in \cite{7}. Another debatable point is the belief
that quantum mechanics (QM) is fundamental. An interesting notion that has been discussed
in the literature is that of a causal set \cite{15}: at smallest scales a locally finite set of elements
endowed with a partial order is postulated. I find the ideas like uniformly embedded points,
and the assumption on the 4-volume in this approach interesting. However, in our approach
space is a continuum and isolated objects correspond to an approximation, while time is the
cause itself, and postulated to be discrete. As regards to the definition of a partially ordered
set that is provided an order relation that is transitive and noncircular \cite{15}, this is nothing but
a restatement of time’s arrow. Rather provocative claim by Barbour is that time is an illusion \cite{16}. A careful study of relativity (both special and general) shows that time is merely a
parameter in them that labels changes in spatial relations \cite{11}, therefore, it is not surprising to arrive at
the conclusion that time does not exist at all by Barbour. However, Barbour’s approach
could be criticized on several ground, see \cite{17}. In my opinion, the emphasis on QM in any
fundamental approach to space and time is most probably flawed: in spite of recent claims
demonstrating weirdness of QM, there are many unanswered questions related with single
system and QM \cite{18,19}.

\section{\bf Conclusion}

Anisotropy in the CMB radiation observed by COBE DMR, and by BOOMERANG and MAXIMA experiments had been pointed out in Chapter 8 in \cite{10}. More precise acoustic peaks in the anisotropies of CMB measured by Planck satellite launched in 2009 provide significant information \cite{20,21}. The first peak indicates that universe is flat, and dark energy is $\Omega_\Lambda = 0.69$. Relative height between the first and the second peaks shows that the amount of baryonic matter is $5\%$ of critical density. It is clear that the question of dark energy is of great importance and current interest.

In conclusion, the ideas put forward here have immense potential to address some of
the challenging fundamental questions in physics and cosmology. Admittedly the progress of
this programme based on the space-time interaction hypothesis has been slow, but considering
the fact that it has so far been a solo effort of the author, it is not too bad either: the present
approach articulates a new paradigm for unity in physics \cite{22}.

Acknowledgement:
The Library facility of Banaras Hindu University, Varanasi is acknowledged.


\begin{thebibliography}{99}
\bibitem{1} A.D. Lewis et al, Ap. J. 586, 135 (2003)
\bibitem{2} S. Perlmutter, Phys. Today 56(4), 53 (2003)
\bibitem{3} R.A. Knop et al, arXiv: astro-ph/0309368
\bibitem{4} P.J.E. Peebles and B. Ratra, Rev. Mod. Phys. 75, 559 (2003)
\bibitem{5} M. Rees, Phil. Trans. Roy Soc. London, 361, 2427 (2003)
\bibitem{6} S. Carlip, Rep. Prog. Phys., 64, 885 (2001)
\bibitem{7} R. Penrose, in Quantum Theory and Beyond, Ed. By Ted Bastin (C.U.P. 1971) pp.151-180
\bibitem{8} D. Oriti, Re. Prog. Phys., 64, 1703 (2001)
\bibitem{9} S.C. Tiwari, in Proc. Einstein Centenary Symp., Ed. By K. Kondo and T.M. Karade (1980) Vol. 1, pp. 303-307: arXiv: gr-qc/0208068
\bibitem{10} S.C. Tiwari, Superluminal Phenomena in Modern Perspective: Faster than Light Signals: Myth or Reality? (Rinton Press, N.J. 2003)
\bibitem{11} S.C. Tiwari, Reality of Time, Chapter 12 in Instantaneous Action at a Distance in Modern Physics: Pro and Contra, Ed. by A. E. Chubykalo et al (Nova, N.J. 1999); arXiv: physics/0110022
\bibitem{12} Notes from Bergson and Modern Physics, M. Capek, Ed. by R.S. Cohen and M.W.
Wartosky (Reidel, Dardrecht 1971)
\bibitem{13} S.C. Tiwari, Rebirth of the Electron: Electromagnetism An unorthodox new approach to fundamental problems in physics (Monograph, First published in 1997; lulu.com, 2006); S.C.
Tiwari, J. Mod. Opt., 46, 1721 (1999)
\bibitem{14} M. Ahmad et al., arXiv: astro-ph/0209274
\bibitem{15} L. Bombelli et al, Phys. Rev. Lett., 59, 521 (1987)
\bibitem{16} J.B. Barbour, The End of Time (O.U.P. 1999)
\bibitem{17} B. Carr, Phys. World (Nov. 1999) pp. 37-39.
\bibitem{18} S.C. Tiwari, Phys. Rev. A65, 016101 (2001)
\bibitem{19} S. C. Tiwari, J. of Optics B. Quantum and Semiclassical Optics, 4, S39 (2002)
\bibitem{20} Planck Collaboration, arXiv: 1807.06205
\bibitem{21} Planck Collaboration, arXiv: 1807.06209
\bibitem{22} S. C. Tiwari, Time-Transcendence-Truth, IONP Studies in Natural Philosophy Volume 1 (lulu.com, 2006)
\end{thebibliography}
\end{document}